%% file: racioppi.tex
\documentclass[11pt]{article}
\usepackage{xspace}
\usepackage{graphicx}
\usepackage{amsmath}
\usepackage{amssymb}
\usepackage{color}
\usepackage{amsfonts}
\usepackage{comment}

\input econfmacros.tex

\def\Title#1{\begin{center} {\Large {\bf #1} } \end{center}}

\begin{document}

\Title{Inflation and classical scale invariance}

\bigskip\bigskip

\begin{raggedright}

Antonio Racioppi\index{Racioppi, A.}, {\it National Institute of Chemical Physics and Biophysics, Tallinn, Estonia}\\

\end{raggedright}

{\small
\begin{flushleft}
\emph{To appear in the proceedings of the Interplay between Particle and Astroparticle Physics workshop, 18 -- 22 August, 2014, held at Queen Mary University of London, UK.}
\end{flushleft}
}

\section{Introduction}

The Background Imaging of Cosmic Extragalactic Polarization (BICEP2) measurement of tensor modes from large angle Cosmic Microwave Background (CMB)
B-mode polarisation~\cite{Ade:2014xna} may confirm the last missing generic prediction of
the inflationary paradigm~\cite{Guth:1980zm,Linde:1981mu,Albrecht:1982wi} -- the existence of primordial tensor perturbations from gravity
waves~\cite{Mukhanov:1981xt,Bardeen:1983qw,Mukhanov:1990me}.
BICEP2 claims to have detected primordial gravitational waves, measuring the tensor-to-scalar ratio to be~\cite{Ade:2014xna}
\begin{equation}
r=0.20^{+0.07}_{-0.05}.
\label{r}
\end{equation}
It is still unclear how much such a measurement will be affected by the subtraction of the dust foreground \cite{Mortonson:2014bja,Flauger:2014qra}.
The BICEP2 result, if confirmed, corresponds to the Hubble parameter $H_*\sim 10^{14}$~GeV
and inflaton potential $V \sim (10^{16}~\text{GeV})^4$ during inflation. This implies, in a model-independent way via the Lyth bound~\cite{Lyth:1996im,Boubekeur:2005zm},
trans-Planckian values for the inflaton field. This also implies that the measured primordial curvature perturbations are dominantly generated by a
slowly rolling inflaton, disfavouring curvaton scenarios~\cite{Lyth:2014yya} and allowing to rule in or rule out the slow-roll inflation with future measurements of
non-Gaussianity $f_{\rm{NL}}$~\cite{Lyth:2014yya,Kehagias:2014wza}.
Although alternative scenarios  can be saved by adding extra fields and dynamics to models,
in the following we shall assume that the BICEP2 result will be substantially confirmed and that it favours generic trans-Planckian single field slow-roll inflation.

Perhaps the most intriguing and most studied consequence of the BICEP2 result is the high scale of inflation.
Following the standard Wilsonian prescription, the inflaton $\phi$ potential can be written as
\begin{equation}
V=V_{\rm ren} + \sum_{n=5}^\infty c_n \frac{\phi^n}{M_{\rm P}^{n-4}},
\label{V}
\end{equation}
where $V_{\rm ren}$ is the renormalisable part of inflaton potential, $M_{\rm P}=2.4\times 10^{18}$~GeV is the reduced Planck mass, and $c_n$ are
the Wilson coefficients of gravity-induced higher order operators. Since BICEP2 implies $\phi/M_{\rm P}\gtrsim (1\div 10)$~ \cite{Antusch:2014cpa}, the infinite sum of
non-renormalisable operators in (\ref{V}) is badly divergent, predicting $V \gg (10^{16}~\text{GeV})^4$ and messing up
the inflation~\cite{Calmet:2014lga} and (meta)stability of the scalar potential \cite{Branchina:2013jra}.

 Trans-Planckian inflaton values have created a lot of confusion in physics community. The proposed solutions vary from assumptions that
 the unknown UV theory of gravity is such that all non-renormalisable operators are exponentially suppressed~\cite{Calmet:2014lga,Salvio:2014soa,Kaloper:2014zba,Chialva:2014rla}, to abandoning the inflation as the origin
 of density perturbations \cite{Kehagias:2014wza}.
However, in the light of successful experimental verification
 of all five generic predictions of the slow-roll inflation (almost scale-invariant density perturbations, adiabatic initial conditions, nearly Gaussian fluctuations,
 spatial flatness, and, finally, tensor perturbations from gravity waves) one should first study the implications of the BICEP2 result to our understanding of
 quantum field theories. This is the aim of our work.

 We argue that the apparent absence of the Planck scale induced operators (\ref{V}), as proven experimentally by the BICEP2 result,
 is an evidence for classically scale-free fundamental physics \cite{Kannike:2014mia}. This implies that all scales in physics are generated by quantum effects. We show that this
 paradigm can be extended also to inflation in a phenomenologically successful way.  We present a most minimal scale-free inflation model
 and show that the result is predictive and consistent with the BICEP2 and Planck measurements.

The idea of scale-invariant inflation is not new. Already the very first papers on inflation \cite{Linde:1981mu,Albrecht:1982wi,Linde:1982zj,Ellis:1982ws,Ellis:1982dg}
considered dynamically induced inflaton potentials {\it \`{a}~la} Coleman-Weinberg~\cite{Coleman:1973jx}.
 Since then  the Coleman-Weinberg inflation has been extensively studied in the context of grand unified theories
\cite{Langbein:1993ym,GonzalezDiaz:1986bu,Yokoyama:1998rw,Rehman:2008qs} and in $U(1)_{B-L}$ extension of the
SM~\cite{Barenboim:2013wra,Okada:2013vxa}. The common feature of all those models, probably inherited from the original Coleman-Weinberg paper~\cite{Coleman:1973jx},
is that the dynamics leading to dimensional transmutation is induced by new gauge interactions beyond the SM.
However, the dimensional transmutation does not need extra gauge interactions!
It can occur just due to running of some scalar quartic coupling $\lambda(\mu)\phi^4$ to negative values at some energy scale $\mu$ due to couplings to other scalar fields,
generating non-trivial physical potentials as demonstrated in ~\cite{Heikinheimo:2013fta}. The models of this type are simple and generic, and therefore we
call them the minimal scale-free (or Coleman-Weinberg) models. In this proceeding we study this type of inflation models.

\section{The model}
\subsection{Properties of single field scale-free inflation scenario}
We start by taking a model-independent approach and assume that the shape of the potential is
generated dynamically by one-loop effects without specifying the underlying physics.
One concrete model realisation will be presented in the next subsection.
The tree-level potential to start with is
\begin{equation}
 V =\Lambda^4  + \frac{\lambda_\phi}{4} \phi^4  ,
 \label{V0}
\end{equation}
where $\phi$ is the inflaton field, $M_P = 2.4 \times 10^{18}$~GeV is the reduced Planck mass and $\Lambda$ is the cosmological constant needed to tune
the potential of the minimum to zero. While particle physics observables depend only on the difference of
the potential, gravity couples to the absolute scale creating the cosmological constant problem that so far does not have a commonly accepted elegant solution. In the following we view the existence of $\Lambda$
as a phenomenological necessity and accept the fine tuning associated with it.

In realistic models of inflation one has to consider effects of inflaton couplings to itself and to other fields.
At one loop level, the renormalisation group improved effective inflaton potential becomes
\begin{equation}
 V_{\rm eff} = \Lambda^4 + \frac{\beta_{\lambda_\phi}}{4} \ln \left|\frac{\phi}{\phi_0} \right| \phi^4
 \label{Vrunbeta},
\end{equation}
where the beta-function $\beta_{\lambda_\phi}$ describes running of $\lambda_\phi$ due to inflaton couplings and $\phi_0$ is the
scale induced by dimensional transmutation that is closely related to the minimum of the potential.
Unlike the previous works
\cite{Linde:1981mu,Albrecht:1982wi,Langbein:1993ym,GonzalezDiaz:1986bu,Yokoyama:1998rw,Rehman:2008qs,Barenboim:2013wra}, we will make do \emph{without} extending the gauge symmetries of the SM.
Treating $\beta_{\lambda_\phi}$ as a constant parameter, $V$ has a minimum at
\begin{equation}
  v_\phi = \frac{\phi_0 }{\sqrt[4]{e}} .
  \label{vevphi}
 \end{equation}
According to our assumptions,  the cosmological constant $\Lambda$ is adjusted so that $V(v_\phi)=0$.
This is needed in order to avoid $\phi=v_\phi$ as an allowed inflaton initial configuration that leads to eternal inflation and the concerning issues \cite{Guth:2007ng}.
Solving this constraint for $\Lambda$, we get
\begin{equation}
 \Lambda = \phi_0 \sqrt[4]{\frac{\beta_{\lambda_\phi}}{16 e}}.
 \label{lam}
\end{equation}

\begin{figure}[!t]
\begin{center}
\includegraphics[width=0.4\columnwidth]{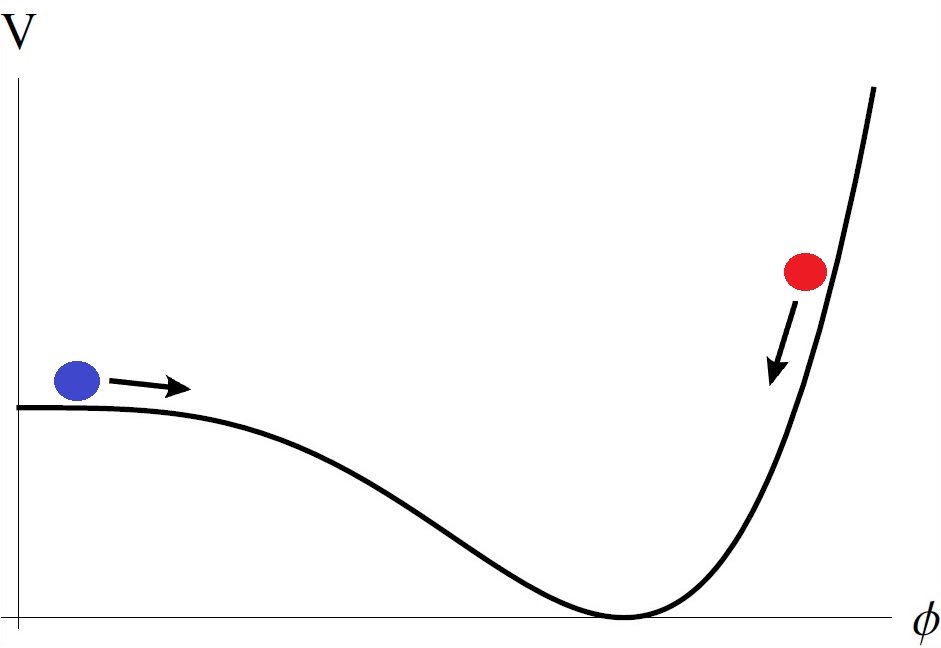}
 \caption{The shape of scale-free inflaton potential. Both chaotic (red) and hilltop (blue) inflation are allowed.}
 \label{fig:Vfig}
\end{center}
\end{figure}

\begin{figure}[!ht]
\begin{center}
\includegraphics[width=0.47\columnwidth]{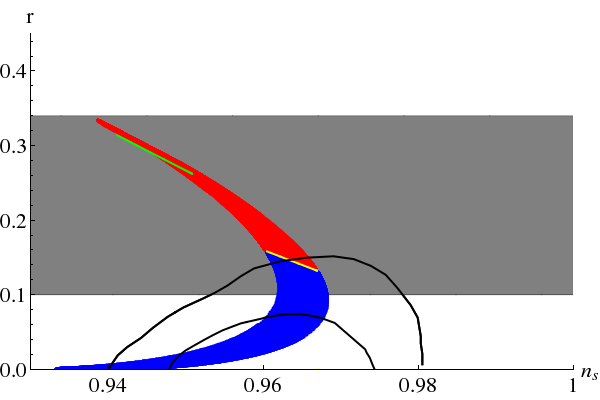}
\includegraphics[width=0.49\columnwidth]{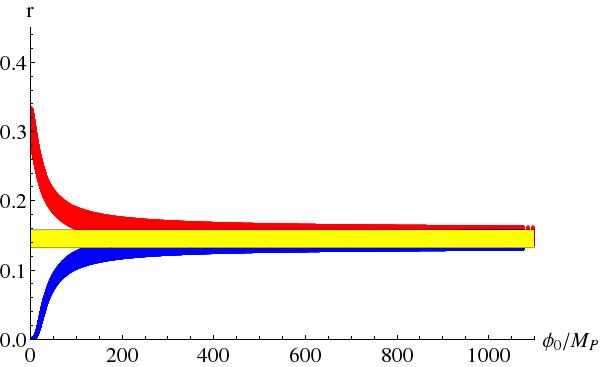}
(a) \hspace{6cm} (b)
 \caption{Predictions for tensor-to-scalar ratio $r$ as a function of $n_s$ (a)
 and as function of $\phi_0$ (b) for $N \in [50,60]$ $e$-folds.
  The blue region represents the hilltop inflation configuration while the red one the chaotic inflation.
  For reference we also plot predictions of  $\frac{m^2}{2} \phi^2$ (yellow) and $ \lambda \phi^4$ (green) potentials.
The $2\sigma$ BICEP2 band (gray) and $2\sigma,$ $1\sigma$ Planck bounds (black) are also presented.
}
 \label{fig:rvsn}
\end{center}
\end{figure}

The potential in eq. (\ref{Vrunbeta}) represents a dynamical realization of the inflaton potential.
The shape of the potential is illustrated in Fig.~\ref{fig:Vfig}.
Such a shape allows for two different, generic types of inflation depending on the initial conditions:
\begin{itemize}
 \item[i.] Small-field hilltop inflation, when $\phi$ rolls down from small field values towards $v_\phi$
 \item[ii.] Large-field chaotic inflation, when $\phi$ rolls down from large field values  towards $v_\phi$.
\end{itemize}
Following~\cite{Kinney:2009vz},  it is straightforward to compute the slow roll parameters, number of $e$-folds $N$, spectral index $n_s$ and its scale dependence,
tensor-to-scalar ratio $r$ and other inflation observables for the potential (\ref{Vrunbeta}). We study for which parameter space this potential
can support phenomenologically acceptable inflation.
The result in the $(n_s,r)$ plane is presented in Fig.~\ref{fig:rvsn}a.
The red and the blue regions correspond to the predictions of our model producing $N \in [50,60]$ $e$-folds of inflation.
We considered $\phi_0$ in the range $[0.1,1000] \, M_P$.
The blue region represents the hilltop inflation configuration while the red one corresponds to the chaotic inflation.
For reference and for interpretation of our result we also plot the predictions of
$V =  \frac{m^2}{2} \phi^2$ (yellow) and    $V =  \lambda \phi^4$ (green) potentials.
The hilltop inflation takes place in the region under the yellow line, while chaotic inflation occurs in the region above  it.
The chaotic region, of course, also contains the simple $\lambda \phi^4$ model.
The  grey band represents the $2\sigma$ BICEP2 result while the black lines are the $1\sigma$ and $2\sigma$ Planck bounds.

It follows from Fig.~\ref{fig:rvsn}a that the spectral index and the tensor-to-scalar ratio are strongly correlated in the considered scale-free inflation scenario.
Present experimental accuracy allows for quite large model parameter space
consistent with the BICEP2 result, with the Planck result, and with the both.
After BICEP2, the $\frac{m^2}{2} \phi^2$ chaotic inflation potential has got a lot of attention since its predictions
agree well with experimental results.
In our case this corresponds to the limit of very large inflaton vacuum expectation value (VEV), $v_\phi \gg M_P.$
In this limit the shape of inflaton potential around the minimum becomes symmetric and inflation observables
lose sensitivity to the initial conditions. To explain this limit better, we plot in Fig. \ref{fig:rvsn}b the predicted range for $r$ in function of $\phi_0$
producing $N \in [50,60]$ $e$-folds of inflation in our model and in the $\frac{m^2}{2} \phi^2$ inflation. The colour code is the same as in previous figures.
We can see that for $v_\phi \gg M_P$, the three different regions overlap.
Therefore, if future data will determine  $(n_s,r)$ along the yellow line in Fig.~\ref{fig:rvsn}a, this will
support the scale-free inflation with a trans-Planckian inflaton VEV.
We note that the parameter space considered in Ref.~\cite{Barenboim:2013wra} corresponds to the tail of blue region in Fig.~\ref{fig:rvsn}a
with $r\ll 0.1$ and $n_s< 0.945.$ Therefore those authors mistakenly concluded  that the scale-free inflation is not consistent with Planck results.
The reason for that is that they considered only inflaton field values below the Planck scale. In our opinion this assumption is overly restrictive.

\subsection{The minimal scale-free model for inflation}
In this subsection we present the minimal scale-free inflation model giving rise to the inflaton potential (\ref{Vrunbeta}).
We consider the following Lagrangian which includes two real singlet scalar fields $\phi$ and $\eta$ and three heavy singlet right-handed neutrinos $N_{i}$:
\begin{align}
 {\cal L} &\supset \frac{1}{2}\partial_\mu\phi \partial^\mu\phi +  \frac{1}{2}\partial_\mu\eta \partial^\mu\eta + {\cal L}_Y - V, \\
  {\cal L}_Y &= Y_\phi^{ij} \bar N^c_{i }N_{j} \phi  +  Y_\eta^{ij} \bar N^c_{i }N_{j} \eta,  \\
 V& = \Lambda^4 + \frac{\lambda_\phi}{4} \phi^4 + \frac{\lambda_{\phi \eta}}{4} \eta^2 \phi^2 + \frac{\lambda_\eta}{4} \eta^4, \label{V3}
\end{align}

where $ {\cal L}_Y$ presents scalar Yukawa couplings involving the right-handed neutrinos needed for the reheating of the Universe.
We assume that the kinetic terms are canonically normalised and there are no explicit mass terms in the scalar potential $V$.
The couplings in  ${\cal L}_Y$ run according to the coupled set of renormalisation group equations (RGEs) given in the Appendix of \cite{Kannike:2014mia}.

The one-loop inflaton effective potential can be written as
\begin{equation}
V_{\rm eff }= V + \Delta V,
\end{equation}
where the loop-level contribution reads
\begin{equation}
\Delta V = \frac{1}{64 \pi^2} \left[ \sum_{i=1}^2 m_i^4 \left(\ln \frac{m_i^2}{\mu^2} - \frac{3}{2} \right)
 - 2 \tr{\left \{ M_{N} M_{N}^\dagger \left(\ln \frac{M_{N} M_{N}^\dagger }{\mu^2} - \frac{3}{2} \right) \right\} } \right].
\end{equation}
Here $m_{i}^2$ are the eigenvalues of the field-dependent scalar mass matrices
\begin{equation}
m_{\phi\eta}^2 =
\begin{pmatrix}
3 \lambda_\phi \phi^2 + \frac{1}{2} \lambda_{\phi\eta} \eta^2 & \lambda_{\phi\eta} \phi \eta \\
 \lambda_{\phi\eta} \phi \eta & 3 \lambda_\eta \eta^2 + \frac{1}{2} \lambda_{\phi\eta} \phi^2
\end{pmatrix},
\end{equation}
\begin{equation}
M_N = Y_\phi \phi + Y_\eta \eta,
\end{equation}
and $\mu$ is the renormalisation scale. Inflation will take place in the direction $\eta=0$, which is the
minimal value  for the field $\eta$. We will see later that such an
assumption is self-consistent. The RGE improved effective
potential for the direction of $\phi$ reads
\begin{align}
V_{\rm eff} &=\Lambda ^4 + \frac{ \lambda _{\phi } (\mu) \phi^4}{4}
     +\frac{  \lambda _{\phi \eta
   }^2} {256 \pi ^2}    \left(\ln \frac{\phi ^2 \lambda _{\phi \eta }}{2 \mu
   ^2}-\frac{3}{2}\right)    \phi ^4, \nonumber
\end{align}
where we neglect the  one loop contribution of $\lambda_\phi$ (since it will be extremely small \cite{Kannike:2014mia})
and of the heavy neutrinos. The beta function $\beta_{\lambda_\phi}$ is dominated by $\lambda_{\phi \eta}$.
In the previous section we treated $\beta_{\lambda_\phi}$ as a constant. It can be easily checked that also this assumption is consistent.
 Therefore
\begin{equation}
 V_{\rm eff} =\Lambda ^4 + \frac{ \lambda _{\phi \eta }^2 }{{256 \pi ^2}}
 \left(\ln \frac{\mu^{2} }{\mu _0^{2}} + \ln \frac{ \lambda _{\phi \eta } \phi^2}{2 \mu ^2}-\frac{3}{2} \right) \phi^4,
\end{equation}
where $\mu_{0}$ is defined by $\lambda_{\phi}(\mu_{0}) = 0$.
We can eliminate the dependence on the renormalisation scale, getting
\begin{equation}
  V_{\rm eff} =\Lambda ^4+\frac{\lambda_{\phi \eta} ^2  \ln \frac{\phi^{2}
   }{\phi_0^{2}}}{256 \pi ^2} \phi ^4,
\end{equation}
where $\phi_0=\sqrt{\frac{2 e^{3/2}}{\lambda _{\phi \eta }}} \mu _0 $. We
see that we have reproduced the potential in eq.~(\ref{Vrunbeta}).
Therefore, the results presented in the previous
section all hold for this model, up to a proper redefinition of the
parameters.

To conclude we give some numerically details about the other parameters involved. In the allowed hilltop region $\lambda(\phi^*) \in [-2,0] \times 10^{-14}$
while in the chaotic one  $\lambda(\phi^*) \in [-3,0] \times 10^{-16}$, where $\phi^*$ is the field value at the ``beginning'' of inflation \cite{Kinney:2009vz}. The portal $\lambda_{\phi\eta} \lesssim 10^{-5}$, therefore it is
 small enough to be treated as a constant, and since $\beta_{\lambda_\phi} \propto \lambda_{\phi\eta}^2$, also $\beta_{\lambda_\phi}$ can be treated as a constant as well. The inflaton mass is $m_\phi \sim 10^{13}$ GeV, while $ m_\eta \sim 10^{17}$~GeV.
 Therefore $m_\eta$ is bigger than the inflation scale $V^*\sim 10^{16}$~GeV, and much bigger than the inflaton mass $m_\phi$ therefore $\eta$ is decoupled and is frozen at $\eta=0$ during inflation\footnote{It has been shown in Ref.~\cite{Lebedev:2012sy}
that $\eta=0$ is quickly achieved during inflation due to the coupling $\lambda_{\phi \eta}.$}.
Thus the model is self-consistent. In order to allow the inflaton decay into right-handed neutrino we must have $m_\phi > 2 m_N $. It can be easily checked that this implies that $Y_\phi$ is neglectable in RGE and that the reheating temperature $T_{RH} \sim 10^7$ GeV.
Finally, for consistency, $\lambda_{\eta} > 0$, but otherwise can have any value that do not spoil perturbativity.

\section{Summary}

The BICEP2 measurement (\ref{r}), that we assume will be substantially confirmed by other experiments, motivated us to study predictions of classically scale-free single field inflation.
In this scenario the inflaton potential as well as all mass scales are generated dynamically
via dimensional transmutation due to inflaton couplings to other fields.
 Since the inflaton field must take trans-Planckian values, gravity above the Planck scale must couple weakly to particle physics.
 Classical scale invariance provides also a natural solution to the absence of large Planck suppressed non-renormalisable operators.
Such classically scale-free models of gravity have been proposed recently in Ref.~\cite{Salvio:2014soa}.

First, working model independently with Coleman-Weinberg type single inflaton potential (\ref{Vrunbeta}), we computed which values of $r$ and $n_s$
the model can accommodate. The results show that $r$ and $n_s$ are strongly correlated but the present experimental accuracy does not allow to specify the model parameters,
and almost any value of $r$ is, in principle, achievable. However, if future measurements will determine $r$ and $n_s$ with high accuracy, this
scenario must pass non-trivial tests. Interestingly, if the future result is consistent with the prediction of  tree level potential $m^2\phi^2,$
in the scale-free inflaton scenario this corresponds to very large scale inflaton physics with its VEV above the Planck scale.

We have presented a minimal scale-free inflaton model and shown with explicit computation how the inflaton potential arises
from dimensional transmutation. We conclude that
classically scale-free inflation models are attractive, self-consistent framework to address physics above Planck scale.

\bigskip
\section{Acknowledgments}

The author thanks K. Kannike and M. Raidal for useful discussions and collaboration.
This work was supported by  grants MJD298,  MTT60,  IUT23-6, CERN+, and by EU through the ERDF  CoE program.

\bibliographystyle{h-physrev}
\bibliography{citations}

\end{document}

%% file: econfmacros.tex
\definecolor{Red}{rgb}{1,0,0}
\definecolor{Green}{rgb}{0,1,0}
\definecolor{Blue}{rgb}{0,0,1}
\definecolor{Black}{rgb}{0,0,0}

\def\beq{\begin{equation}}
\def\eeq#1{\label{#1}\end{equation}}
\def\eeqn{\end{equation}}

\def\beqa{\begin{eqnarray}}
\def\eeqa#1{\label{#1}\end{eqnarray}}
\def\eeqan{\end{eqnarray}}

\let\bar=\overbar

\def\tr{{\mbox{\rm tr}}}

\def\Dslash{\not{\hbox{\kern-4pt $D$}}}
\def\dslash{\not{\hbox{\kern-2pt $\del$}}}

\def\msb{{\bar{\ssstyle M \kern -1pt S}}}

%% file: racioppi.bbl
\begin{thebibliography}{10}

\bibitem{Ade:2014xna}
BICEP2 Collaboration, P.~Ade {\em et~al.},
\newblock (2014), 1403.3985.

\bibitem{Guth:1980zm}
A.~H. Guth,
\newblock Phys.Rev. {\bf D23}, 347 (1981).

\bibitem{Linde:1981mu}
A.~D. Linde,
\newblock Phys.Lett. {\bf B108}, 389 (1982).

\bibitem{Albrecht:1982wi}
A.~Albrecht and P.~J. Steinhardt,
\newblock Phys.Rev.Lett. {\bf 48}, 1220 (1982).

\bibitem{Mukhanov:1981xt}
V.~F. Mukhanov and G.~V. Chibisov,
\newblock JETP Lett. {\bf 33}, 532 (1981).

\bibitem{Bardeen:1983qw}
J.~M. Bardeen, P.~J. Steinhardt, and M.~S. Turner,
\newblock Phys.Rev. {\bf D28}, 679 (1983).

\bibitem{Mukhanov:1990me}
V.~F. Mukhanov, H.~Feldman, and R.~H. Brandenberger,
\newblock Phys.Rept. {\bf 215}, 203 (1992).

\bibitem{Mortonson:2014bja}
M.~J. Mortonson and U.~Seljak,
\newblock JCAP {\bf 1410}, 035 (2014), 1405.5857.

\bibitem{Flauger:2014qra}
R.~Flauger, J.~C. Hill, and D.~N. Spergel,
\newblock JCAP {\bf 1408}, 039 (2014), 1405.7351.

\bibitem{Lyth:1996im}
D.~H. Lyth,
\newblock Phys.Rev.Lett. {\bf 78}, 1861 (1997), hep-ph/9606387.

\bibitem{Boubekeur:2005zm}
L.~Boubekeur and D.~Lyth,
\newblock JCAP {\bf 0507}, 010 (2005), hep-ph/0502047.

\bibitem{Lyth:2014yya}
D.~H. Lyth,
\newblock (2014), 1403.7323.

\bibitem{Kehagias:2014wza}
A.~Kehagias and A.~Riotto,
\newblock (2014), 1403.4811.

\bibitem{Antusch:2014cpa}
S.~Antusch and D.~Nolde,
\newblock (2014), 1404.1821.

\bibitem{Calmet:2014lga}
X.~Calmet and V.~Sanz,
\newblock (2014), 1403.5100.

\bibitem{Branchina:2013jra}
V.~Branchina and E.~Messina,
\newblock Phys.Rev.Lett. {\bf 111}, 241801 (2013), 1307.5193.

\bibitem{Salvio:2014soa}
A.~Salvio and A.~Strumia,
\newblock (2014), 1403.4226.

\bibitem{Kaloper:2014zba}
N.~Kaloper and A.~Lawrence,
\newblock (2014), 1404.2912.

\bibitem{Chialva:2014rla}
D.~Chialva and A.~Mazumdar,
\newblock (2014), 1405.0513.

\bibitem{Kannike:2014mia}
K.~Kannike, A.~Racioppi, and M.~Raidal,
\newblock JHEP {\bf 1406}, 154 (2014), 1405.3987.

\bibitem{Linde:1982zj}
A.~D. Linde,
\newblock Phys.Lett. {\bf B114}, 431 (1982).

\bibitem{Ellis:1982ws}
J.~R. Ellis, D.~V. Nanopoulos, K.~A. Olive, and K.~Tamvakis,
\newblock Nucl.Phys. {\bf B221}, 524 (1983).

\bibitem{Ellis:1982dg}
J.~R. Ellis, D.~V. Nanopoulos, K.~A. Olive, and K.~Tamvakis,
\newblock Phys.Lett. {\bf B120}, 331 (1983).

\bibitem{Coleman:1973jx}
S.~R. Coleman and E.~J. Weinberg,
\newblock Phys.Rev. {\bf D7}, 1888 (1973).

\bibitem{Langbein:1993ym}
R.~Langbein, K.~Langfeld, H.~Reinhardt, and L.~von Smekal,
\newblock Mod.Phys.Lett. {\bf A11}, 631 (1996), hep-ph/9310335.

\bibitem{GonzalezDiaz:1986bu}
P.~Gonzalez-Diaz,
\newblock Phys.Lett. {\bf B176}, 29 (1986).

\bibitem{Yokoyama:1998rw}
J.~Yokoyama,
\newblock Phys.Rev. {\bf D59}, 107303 (1999).

\bibitem{Rehman:2008qs}
M.~U. Rehman, Q.~Shafi, and J.~R. Wickman,
\newblock Phys.Rev. {\bf D78}, 123516 (2008), 0810.3625.

\bibitem{Barenboim:2013wra}
G.~Barenboim, E.~J. Chun, and H.~M. Lee,
\newblock Phys.Lett. {\bf B730}, 81 (2014), 1309.1695.

\bibitem{Okada:2013vxa}
N.~Okada and Q.~Shafi,
\newblock (2013), 1311.0921.

\bibitem{Heikinheimo:2013fta}
M.~Heikinheimo, A.~Racioppi, M.~Raidal, C.~Spethmann, and K.~Tuominen,
\newblock Mod.Phys.Lett. {\bf A29}, 1450077 (2014), 1304.7006.

\bibitem{Guth:2007ng}
A.~H. Guth,
\newblock J.Phys. {\bf A40}, 6811 (2007), hep-th/0702178.

\bibitem{Kinney:2009vz}
W.~H. {Kinney},
\newblock ArXiv e-prints  (2009), 0902.1529.

\bibitem{Lebedev:2012sy}
O.~Lebedev and A.~Westphal,
\newblock Phys.Lett. {\bf B719}, 415 (2013), 1210.6987.

\end{thebibliography}
